\documentstyle[preprint,aps,epsfig]{revtex}
\begin{document}
\preprint{\vbox{\hbox{IFT--P.024/97}\hbox{hep-ph/9703346}}}
\title{Bounds on Contact Interactions from LEP1 Data \\
and the High--\boldmath{$Q^2$} HERA Events} 
\author{M.\ C.\ Gonzalez--Garcia$^{1,2}$   and S.\ F.\ Novaes$^1$}
\address{$^1$ Instituto de F\'{\i}sica Te\'orica, 
Universidade  Estadual Paulista, \\  
Rua Pamplona 145, CEP 01405--900 S\~ao Paulo, Brazil.\\
and \\
$^2$ Instituto de F\'{\i}sica Corpuscular -- IFIC/CSIC \\
Dept.\ de F\'{\i}sica Te\'orica, Universidad de Valencia \\
46100 Burjassot, Valencia, Spain}
\date{\today}
\maketitle
\widetext
\begin{abstract}
A contact four--fermion interaction between light quarks and
electrons has been evoked as a possible explanation for the
excess of events observed by HERA at high--$Q^2$. We explore the 
1--loop effects of such interaction in $\Gamma(Z^0 \to e^+ e^-)$
measured at LEP and impose strong bounds on the lower limit of
the effective scale. Our results are able to discard some of the 
contact interactions as possible explanation for the HERA events.
\end{abstract}
\pacs{}

Recently the H1 \cite{h1} and ZEUS \cite{zeus} experiments at
HERA have reported the observation of an excess of events, compared
with the Standard Model prediction, in the reaction $e^+ p \to
e^+ + X$ at very high--$Q^2$. The H1 Collaboration observed
events seem to be concentrated at an invariant mass of $\sim
200$ GeV, what could suggest the presence of a $s$--channel
resonant state.  The ZEUS Collaboration data, however, are more
spread in invariant mass.  The probability of a statistical
fluctuation seems to be quite small (less than $6 \times
10^{-3}$, for the H1 data). Nevertheless, up to this moment, it
is not possible to establish the resonant or continuum aspect of
the events. 

It seems very hard to find an explanation for these events in the
scope of the Standard Model, {\it e.g.\/} modifying the partonic
distribution functions, or including new QCD corrections. Among
the possible new physics explanations for these events, there is
the $s$--channel production of leptoquarks or  squarks of a
R--parity violating supersymmetric model \cite{squark,alt}.
Besides this  scenario, we can think of a non--resonant
interpretation of the HERA data, which involves an effective
four--fermion interaction $eeqq$, where $q = u, d$ quarks
\cite{alt}.

A convenient parametrization of the four--fermion contact
interaction is \cite{peskin},
\begin{equation}
{\cal L}_{eeqq} = g^2 \sum_{i,j = L, R} \sum_{q = u, d} 
\eta_{i,j} \frac{1}{(\Lambda_{ij}^{\eta q})^2} ( \bar{e}_i \gamma^\mu e_i )
(\bar{q}_j \gamma_\mu q_j ) \; ,
\label{lag}
\end{equation}
where $i,j$ refer to the different fermion helicities, and
$\eta_{i,j} = \pm 1$ enables us to consider constructive and
destructive interference with standard contribution to the
processes. Such effective interaction can be generated at low energy
by the exchange of a heavy particle in the $t$--channel between
the quark and the electron lines. This appears naturally in
models where quarks and leptons are composite particles through the
exchange of some common constituent or of the binding particles. In
the same fashion, interaction (\ref{lag}) can be used to describe
the low energy limit of the exchange of a new heavy neutral
particle, like the $Z^\prime$ gauge boson.

In general, bounds on the scale $\Lambda_{ij}^{\eta q}$ are
obtained assuming $g^2/4 \pi = 1$ for the new strong interaction
coupling. Lagrangian (\ref{lag}) has been used in Ref.\
\cite{alt,bar} to fit the integrated $Q^2$ distributions of the
HERA data, taking into account bounds on the scale
$\Lambda_{ij}^{\eta q}$ from CDF Collaboration \cite{cdf} at Tevatron
collider, as well as those from LEP \cite{lep:4}, including the
new ones obtained by the OPAL Collaboration at $\sqrt{s}=170$,
$172$ GeV \cite{koma}. Altarelli {\it et al.} best fits were
obtained for the $RL$ or $LR$ polarizations with the minimum
allowed value for the scale $\Lambda_{ij}^{\eta q}$.

\begin{figure}
\begin{center}
\mbox{\epsfig{file=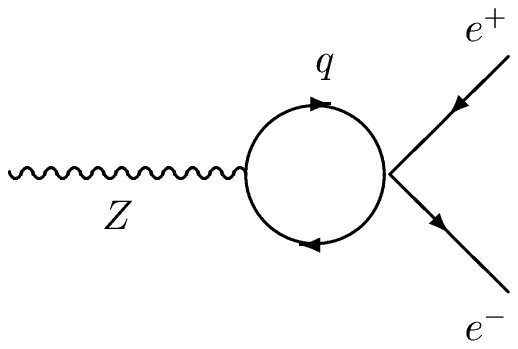,width=1.2\linewidth,height=0.32\textheight,angle=0}}
\end{center}
\caption{Feynman diagram leading to the correction of $\Gamma_{ll}$.}
\label{diagram}
\end{figure}

In this letter, we analyze the one--loop effect of the
interaction (\ref{lag}) in the leptonic width of $Z^0$, and we
employ the most recent LEP data \cite{lep} on $\Gamma(Z^0 \to e^+
e^-)$ to establish strong bounds on the scale
$\Lambda_{ij}^{\eta q}$. We evaluate the relevant Feynman diagram
(see Fig.~\ref{diagram}) in dimensional regularization neglecting
the external (electron) and internal (light quark) fermion
masses.  We retain only the leading non-analytical contributions
from the loop diagram by making the identification
\begin{equation}
\frac{2}{4-d} \rightarrow {\rm{log}}\;\frac{\Lambda^2}{\mu^2}\; ,
\nonumber
\end{equation}
where $d = 4 - 2 \epsilon$ is the space--time dimension,
$\Lambda$ is the energy scale which characterizes the appearance
of new physics, and $\mu$ is the scale involved in the process,
which we choose $\mu=M_Z$ and we drop finite terms.

In this way, we obtain a quite compact result for the light quark loop
contribution of the four--fermion interaction to $\Gamma(Z^0 \to
e^+ e^-) \equiv \Gamma_{ee}$,
\begin{equation}
\Delta \Gamma_{ee} = - \eta_{ij} \frac{\alpha}{6 \pi s_W^2 c_W^2} 
G_i^e G_j^q \frac{M_Z^3}{(\Lambda_{ij}^{\eta q})^2} 
\log \frac{(\Lambda_{ij}^{\eta q})^2}{M_Z^2} \; ,
\label{wid}
\end{equation}
where $s_W  (c_W) = \sin\theta_W (\cos\theta_W)$ and $G_R^f = -
Q^f s_W^2$,  $G_L^f = T_3^f - Q^f s_W^2$, with $T_3^f$, and $Q^f$
being the third component of the weak isospin and electric charge
of the fermion, respectively.

The most recent LEP experimental result \cite{lep} can be
compared with the Standard Model predictions for the leptonic
width, $\Gamma_{ll} = 83.91 \pm 0.11$ TeV, in order to establish
bounds on the scale $\Lambda_{ij}^{\eta q}$ through Eq.\ (\ref{wid}).
The Standard Model result depends on the top quark and Higgs
boson masses and we have generated using ZFITTER \cite{zfit} the
results for $\Gamma_{ll}$ with the top quark mass in the range
$m_{\text{top}}=175 \pm 6 $ GeV and for the Higgs boson mass $M_H=60$,
$300$, and $1000$ GeV (see Table \ref{sm}).

Our limits on the scale $\Lambda_{ij}^{\eta q}$ are summarized in Table
\ref{limits}. We present the 95\% CL lower limit on the scale
$\Lambda_{ij}^{\eta q}$ for different values of $m_{\text{top}}$ and
$M_H$. Some comments are in order. As can be seen from Table
\ref{sm},  the experimental result coincides precisely with the
SM prediction for $m_{\text{top}}=175$ GeV and $M_H=300$ GeV. The
SM expectation is lower (higher) than the measured value for
lighter (heavier) top quark and  heavier (lighter) Higgs boson. In
consequence, those interactions which yield a positive increase in the
leptonic width are more severely constrained for larger
$m_{\text{top}}$ and smaller $M_{H}$. The opposite holds for
interactions which tend to decrease the leptonic width. 
In particular, contact interactions which decrease the leptonic
$Z$ width are ruled out for a heavy Higgs boson and a light
top quark for any value of the scale. 

Table \ref{limits} shows that, taking for instance $m_{\text{top}}
= 175$ GeV, and $M_{H} = 300$ GeV, our limits for 
$\Lambda^{\pm q}_{LL}$, 
$\Lambda^{\pm q}_{RL}$ ($q = u,d$), 
$\Lambda^{\pm u}_{LR}$, and 
$\Lambda^{+ u}_{RR}$
are always stronger than those obtained recently by the OPAL
Collaboration \cite{koma}. In particular for $\Lambda^{+d}_{RL}$ 
our limits are stronger than OPAL bounds for any value of $M_H$
and $m_{\text{top}}$.  This result strongly disfavours the
contact four--fermion interaction term $\Lambda^{+d}_{RL}$ as a
possible solution for the HERA data puzzle. Moreover, other
configurations suggested in Ref.\ \cite{alt,bar}, such as
$\Lambda^{-d}_{LR}=1.7$ TeV, $\Lambda^{+u}_{LR}=2.5$ TeV,
$\Lambda^{+u}_{RL}=2.5$ TeV, or the combination
$\Lambda^{+u}_{LR}=\Lambda^{+u}_{RL} = 3$ TeV are not allowed for
large values of $M_H$ with a light top quark. 

In conclusion, we have shown that the one--loop contribution to
leptonic $Z^0$ width coming from contact effective interactions
involving electrons and light quarks can lead to a strong bound
on the compositeness scale $\Lambda^{\pm q}_{ij}$. These bounds
are in general more stringent than the ones obtained from the
tree--level contribution to the total cross section $e^+ e^- \to
q\bar{q}$  directly measure at LEP \cite{lep:4,koma}.

\acknowledgments
M.\ C.\ Gonzalez--Garcia is grateful to the Instituto de F\'{\i}sica
Te\'orica of the Universidade  Estadual Paulista for its kind
hospitality. This work was supported by Funda\c{c}\~ao de Amparo
\`a Pesquisa do Estado de S\~ao Paulo,by DGICYT under grant
PB95-1077, by CICYT under  grant AEN96-1718,  and by
Conselho Nacional de Desenvolvimento Cient\'{\i}fico e
Tecnol\'ogico.

\protect

\widetext
\begin{table}
\caption{Standard Model prediction for $\Gamma_{ll}$, in MeV,
for different values of $m_{\text{top}}$ and $M_H$.}
\label{sm}
\begin{displaymath}
\begin{array}{|c c|c|c|c|}
\hline
\multicolumn{2}{|c}{  }  & \multicolumn{3}{|c|}{m_{\text{top}}}\\ [-0.2cm]
 \multicolumn{2}{|c}{  } &\multicolumn{1}{|c}{169}  &\multicolumn{1}{c}{175}  
 & \multicolumn{1}{c|}{181}  \\\hline
    & 1000 & 83.72 & 83.77 & 83.82 \\\cline{3-5} 
M_H & 300  & 83.86 & 83.92 & 83.96 \\\cline{3-5}
    &  60  & 83.96 & 84.02 & 84.08  \\
\hline
\end{array}
\end{displaymath}
\end{table} 

\widetext
\begin{table}
\caption{95\% CL limits on the effective contact interaction scale
$\Lambda_{ij}^{\eta q}$ in TeV.  In the entries marked as ``---"no value
of $\Lambda_{ij}^{\eta q}$ is allowed. }
\label{limits}
\begin{displaymath}
\begin{array}{|l|ll|lll||lll||| lll||lll|}
\hline 
\multicolumn{3} {|c|}{ } &
\multicolumn{6}{c|||}{\bar e e \bar u u} & 
\multicolumn{6}{|c|}{\bar e e\bar d d} \\
\hline
   &     &      &  \multicolumn{2}{c}{\eta=-1} &     &  
   \multicolumn{2}{c}{\eta=+1} &     
                &     \multicolumn{2}{c}{\eta=-1} 
                 &     &       \multicolumn{2}{c}{\eta=+1} &     \\
\hline
   &     &      &      &  m_{\text{top}} &     &    & m_{\text{top}}   &      
                &      &  m_{\text{top}} &     &    & m_{\text{top}}   &         \\
   &     &      & 169  &  175 & 181 & 169 & 175  & 181 
                & 169  &  175 & 181 & 169 & 175  & 181 \\
\hline
   &     & 1000 & \mbox{---} &  5.8     & 3.6 & 1.6 & 1.7     & 1.9 
                &   1.8 &   1.9 &   2.2 &  \mbox{---} &   6.5 &   4.1\\
LL & M_H & 300  &  2.9 &   2.4    & 2.1 & 2.1 & 2.4     & 2.9  
                &   2.4 &   2.7 &   3.3 &   3.3 &   2.7 &   2.4\\
   &     & 60   &  2.1 &   1.8  &  1.6  & 2.9 & 4.2    &  12  
                &   3.3 &   4.7  &   14  &   2.4 &   2.1 &   1.8 \\
 \hline
   &     & 1000 &   1.0 &   1.1 &   1.2 &  \mbox{---} &   3.6 &   2.3
                &  \mbox{---}  &   2.4 &   1.5 &   0.6 &   0.7 &   0.8  \\ 
LR & M_H &  300 &   1.3 &   1.5 &   1.8 &   1.8 &   1.5 &   1.3 
                 &   1.2 &   1.0 &   0.8 &   0.8 &   1.0 &   1.2 \\
   &     &  60  &   1.8 &   2.6 &   7.9 &   1.3 &   1.1 &   1.0 
                &   0.8 &   0.7 &   0.6 &   1.2 &   1.7 &   5.3  \\
 \hline
   &     & 1000 &   1.4 &   1.6 &   1.8 &  \mbox{---} &   5.3 &   3.3 
                 &  \mbox{---} &   6.0 &   3.8 &   1.6 &   1.8 &   2.0 \\
RL & M_H &  300 &   1.9  &   2.2 &   2.7 &   2.7 &   2.2 &   1.9 
                &   3.0 &   2.5  &   2.2 &   2.2 &   2.5 &   3.0 \\
   &     &   60 &   2.7  &   3.9 &  11  &   1.9 &   1.7 &   1.5 
                & 2.2 &   1.9 &   1.7 &   3.0 &   4.3 &  13\\
 \hline
   &     & 1000 &  \mbox{---} &   3.4 &   2.1 &   0.9 &   1.0 &   1.1
                &   0.6 &   0.6 &   0.7 &  \mbox{---} &   2.2 &   1.4 \\
RR & M_H &  300 &   1.7 &   1.4 &   1.2 &   1.2 &   1.4 &   1.7 
                &   0.8 &   0.9 &   1.1 &   1.1 &   0.9 &   0.8  \\
   &     &   60 &   1.2 &   1.0 &   0.9 &   1.7 &   2.4 &   7.3 
                &   1.1 &   1.6 &   4.9 &   0.8 &   0.7 &   0.6 \\
 \hline
\end{array}
\end{displaymath}
\end{table}

\end{document}